\documentclass[reprint,showpacs,amssymb,amsmath,aps,prb]{revtex4-1}
\newcommand{\half}{\mbox{$\frac{1}{2}$}}
\usepackage{graphics}
\usepackage{bm}
\usepackage{amsmath}
\usepackage{amssymb}
\begin{document}

\title{Generalized mean field description of entanglement in dimerized spin systems}
\author{A.\ Boette, R.\ Rossignoli, N.\ Canosa, J.\ M.\ Matera}
\affiliation{Departamento de F\'isica-IFLP,
Universidad Nacional de La Plata, C.C. 67, La Plata (1900), Argentina}

\begin{abstract}
We discuss a generalized self-consistent mean field (MF) treatment, based on
the selection of an arbitrary subset of operators for representing the system
density matrix, and its application to the problem of entanglement evaluation
in composite quantum systems. As a specific example, we examine in detail a
pair MF approach to the ground state (GS) of dimerized  spin $1/2$ systems with
anisotropic ferromagnetic-type $XY$ and $XYZ$ couplings in a transverse field,
including chains and arrays with first neighbor and also longer range
couplings. The approach is fully analytic and able to capture the main features
of the GS of these systems, in contrast with the conventional single spin MF.
Its phase diagram differs significantly from that of the latter, exhibiting
($S_z$) parity breaking just in a finite field window if the coupling between
pairs is sufficiently weak, together with a fully dimerized phase below this
window and a partially aligned phase above it. It is then shown that through
symmetry restoration, the approach is able to correctly predict not only the
concurrence of a pair, but also its entanglement with the rest of the chain,
which shows a pronounced peak in the parity breaking window. Perturbative
corrections allow to reproduce more subtle observables like the entanglement
between weakly coupled spins and the low lying energy spectrum. All predictions
are tested against exact results for finite systems.
\end{abstract} 
\maketitle

\section{Introduction}
The analysis of correlations and entanglement in interacting quantum many body
systems has attracted strong attention in recent years \cite{Am.08,ECP.10},
motivated by their deep implications for quantum information processing and
transmission \cite{NC.00}, the impressive advances in techniques for
controlling and measuring quantum systems \cite{HR.06} and the new perspective
they provide for the analysis of quantum phase transitions
\cite{Am.08,ECP.10,ON.02}. While the conventional mean field (MF)
approximations \cite{RS.80} provide a basic starting point for studying such
systems over a broad range of the pertinent control parameters, they are not
directly suitable for the description of entanglement, since they are based on
completely factorized states. More sophisticated treatments have been developed
to include and compute quantum correlations, like for instance density matrix
renormalization group (DMRG) techniques \cite{DMRG.03,OS.95}, matrix product
states and tensor network methods \cite{OS.95,VC.08,GV.08}, variational valence
bond based approximations \cite{LDA.88,LS.07}, quantum Monte Carlo calculations
\cite{VMC.01},  and inclusion of static and quantum fluctuations around MF
\cite{RC.97,CRM.07}. In addition, non-conventional MF approaches, able to
intrinsically include some essential correlations, have also been proposed and
recently improved and revisited \cite{MC.04,DJ.02,Y.09}, which start from the
so-called cluster MF approach, also known as BPW (Bethe-Peierls-Weiss)
approximation \cite{BPW.35}. The essential point in these schemes is the
consideration of composite sites containing more than one ``body''  as the
basic independent units. Their application  to specific spin systems
\cite{MC.04,Y.09} has shown their capability for determining  phase diagrams
and critical temperatures, as well as for describing the main features of
observables such as magnetization and susceptibility. Their ability to predict
entanglement measures has so far not been investigated.

The aim of this work is to investigate a general self-consistent variational MF
treatment, based on the selection of an arbitrary subset of operators for
representing the system density matrix \cite{RP.90}, and its potential for
describing basic entanglement measures in spin systems. The approach can be
applied at both zero or finite temperatures and contains as particular cases
the conventional as well as the cluster-type MF approaches. In contrast with
other variational treatments, the generalized MF scheme  does not require an
explicit ansatz for the approximate GS, as the latter is naturally determined
by the self-consistency relations according to the chosen set of operators. The
scheme may be also used as a convenient starting point for more sophisticated
treatments. We will examine in particular its capability for describing
entanglement, both within the defined units as well as between them, the latter
emerging through symmetry restoration or perturbative corrections.

As a specific example, we will consider a pair MF approximation to the ground
state (GS) of  dimerized spin $1/2$ systems with anisotropic $XY$ or $XYZ$
couplings in a transverse field. In order to test its accuracy, we first
examine the case of dimerized $XY$ chains with first neighbor couplings, where
the exact results for any size  \cite{Pr.75,Pr.77,Pr.09,FM.07,GG.09,CRM.10} can
be obtained through the Jordan-Wigner fermionization \cite{LSM.61}. We then
examine dimerized chains with longer range couplings, dimer lattices and
dimerized $XYZ$ systems, where exact results for finite samples were obtained
by numerical diagonalization. Dimerized systems are of great interest  in both
condensed matter physics and quantum information
\cite{Pr.75,Pr.77,Pr.09,FM.07,GG.09,CRM.10,MG.69,S.05,KSV.07,HXG.08,HH.11,Mn.14},
and can be realized in different ways, including recently cold atoms trapped in
optical lattices \cite{ZZ.14}. Spin $1/2$ systems have the additional advantage of
permitting a direct computation of the pairwise entanglement through the
concurrence \cite{Wo.97}.

While conserving the conceptual simplicity of the conventional MF scheme, we
will show that in contrast with the latter, the pair MF approach is able to
provide a reliable yet still analytic and simple description of dimerized
arrays. Its phase diagram differs significantly from that of the conventional
MF, and clearly identifies, for a wide range of systems,  a fully dimerized
phase for weak fields, a partially aligned phase for strong fields and an
intermediate $S_z$-parity breaking degenerate phase.  It then predicts, in
particular, the two transitions exhibited by the GS of the dimerized $XY$ chain
for increasing fields \cite{Pr.75}, providing a clear approximate picture of
the GS in each phase.  The approach also leads to a reduced pair density which
correctly describes not only the internal entanglement of the pair, but also
(through symmetry restoration) its entanglement with the rest of the system,
which shows a prominent peak precisely in the parity breaking sector. By means
of simple perturbative corrections, the approach can predict the tails of this
entanglement outside the parity breaking sector, as well as the entanglement
between weakly coupled spins and the low lying energy spectrum.  The formalism
is described in sec.\ \ref{II}, while the application to dimerized $XY$ and
$XYZ$ systems is developed in sec.\  \ref{III}, with the exact analytic solution
for the dimerized $XY$ chain discussed in the Appendix. Conclusions are given
in \ref{IV}.

\section{Formalism \label{II}}
\subsection{General self-consistent  approximation}
The mixed state $\rho$ of a system at temperature $T=1/k\beta$ described by a
Hamiltonian $H$, minimizes the free energy functional $F(\rho)=\langle
H\rangle_{\rho}-TS(\rho)$, where $\langle H\rangle_\rho={\rm Tr}\,\rho H$ and
$S(\rho)=-k{\rm Tr}\,\rho\ln\rho$ is the entropy. One can then formulate a
general variational approximation to $\rho$  based on the trial mixed state
\cite{RP.90}
\begin{equation}
\rho_h=\exp[-\beta h]/Z_h,\;\;h=\sum_i\lambda_i O_i\,,\label{h}
\end{equation}
where $Z_h={\rm Tr}\exp[-\beta h]$ and $\{O_i,\;i=1,\ldots,m\}$ is an arbitrary
set of linearly independent operators, with $\lambda_i$ parameters determined
through the minimization of $F(\rho_h)$. Considering the averages $\langle
O_i\rangle\equiv {\rm Tr}\rho_h O_i$, functions of the $\lambda_i$'s, as the
independent parameters, the equations $\frac{\partial F(\rho_h)}{\partial
\langle O_i\rangle}=0$ lead to $\lambda_i=\frac{\partial \langle
H\rangle}{\partial \langle O_i\rangle}$ and hence, to the {\it self-consistent}
approximate Hamiltonian
\begin{equation}
h=\sum_i \frac{\partial \langle H\rangle}{\partial \langle O_i\rangle} O_i\,,
\label{sc}
\end{equation}
where $\langle H\rangle={\rm Tr}\,\rho_h\,H$. If the $O_i$'s form  a complete
set, $H$ is a linear combination of them and Eq.\ (\ref{sc}) leads to $h=H$.
Otherwise, $\langle H\rangle$ will in general be a non-linear function of the
$\langle O_i\rangle's$ and (\ref{h})--(\ref{sc}) lead to a non-linear set of
equations for the $\lambda_i's$. While the basic MF approximations \cite{RS.80}
are obtained when the $O_i$'s are restricted to one-body operators and traces
are taken in the grand canonical ensemble (with $H\rightarrow H-\mu N$), Eq.\
(\ref{sc}) holds for {\it any} restricted set, which may include {\it some}
{two-body} (or in general $n$-body) operators, and for traces taken in {\it
any} subspace ${\cal S}$ invariant under $H$ and all $O_i's$ \cite{RP.90}.

Here we will apply this general scheme to a composite system formed by $N$
distinguishable subsystems, such as an array of spins $\bm{s}_i$ located at
different sites, where the total Hilbert space is $\otimes_{i=1}^N {\cal S}_i$,
with ${\cal S}_i$ that of subsystem $i$. We will consider Hamiltonians
containing local terms and two-body couplings,
\begin{eqnarray}
H&=&\sum_{i}B^i_{\mu}O_{i}^\mu-{\textstyle\frac{1}{2}}\sum_{i\neq j}
 J^{ij}_{\mu\nu}O_{i}^\mu O_{j}^\nu\,, \label{H}\end{eqnarray}
where $O_{i}^\mu$ are local operators pertaining to subsystem $i$
($[O_{i}^\mu,O_j^\nu]=0$ if $i\neq j$) and sum over repeated labels $\mu,\nu$
is implied. The standard MF arises when the $O_i$'s  in (\ref{h})--(\ref{sc})
are restricted to {\it local} operators $O_i^\mu$, i.e., when a ``site'' is
identified with a {\it single} subsystem $i$. The present scheme enables,
however, to consider as well {\it composite} sites $C_k$, such as pairs or
clusters of spins in a spin system, where products $O_i^\mu O_j^\nu$  for sites
$i,j$ in the {\it same} cluster {\it  are also included} within the operators
$O_i$ of (\ref{h})--(\ref{sc}). This is convenient when such pairs or clusters
are internally strongly coupled but interact only weakly between them. The
ensuing self-consistent scheme will treat the internal couplings exactly,
leaving the MF for the weak couplings.

In this approach, $h=\sum_k h_k$, with $h_k$ {\it local in} $C_k$, such that
$\rho_h=\otimes_k \rho_k$, with $\rho_k=\exp[-\beta h_k]/Z_{h_k}$. Hence,
\begin{eqnarray}\langle H\rangle&=&\!\!\!\!\!{\textstyle\sum\limits_{k,i\in C_k}
\!\![B^i_{\mu}\langle O_i^\mu\rangle-
\frac{1}{2}\!\!\!\sum\limits_{j\in C_k}\!\!J^{ij}_{\mu\nu}\langle O_i^\mu O_j^\nu\rangle}
{\textstyle-\frac{1}{2}\!\!\!\sum\limits_{j\notin C_k}\!\!\!J^{ij}_{\mu\nu}
\langle O_i^\mu\rangle\langle O_{j}^{\nu}\rangle]}\,,\nonumber\\
&&\label{6}
\end{eqnarray}
and Eq.\ (\ref{sc}) leads to
\begin{eqnarray}h_k&=&\!\!{\textstyle\sum\limits_{i\in
C_k}[(B^i_{\mu}-\!\!\!\sum\limits_{j\notin C_k}\!\!\!J^{ij}_{\mu\nu}\langle
O_j^\nu\rangle)O_{i}^{\mu}-\frac{1}{2}\!\!\!\sum\limits_{j\in
C_k}\!\!\!J^{ij}_{\mu\nu}O_i^\mu O_j^\nu]}\,,\label{hk}
\end{eqnarray}
 which contains the exact internal two-body terms, as opposed to the standard MF.
 Eq.\ (\ref{hk}) implies the self-consistent conditions
 \begin{equation}\langle O_i^\mu\rangle={\rm Tr}\,\rho_k O_i^\mu\,,\;\;\;i\in C_k
 \,,\label{sc2}
 \end{equation}
to be fulfilled for all $C_k$, which can be solved, for instance,
iteratively, after starting from an initial guess for the
$\langle O_i^\mu\rangle$'s or the associated parameters $\lambda^i_\mu$.
 We will denote  this approach as generalized MF
(GMF). Eq.\ (\ref{H}) can now be rewritten as
\begin{equation}
H=\langle H\rangle+\sum_k[h_k-\langle h_k\rangle-{\textstyle\frac{1}{2}}\!\!\!\!\!\sum_{i\in C_k,j\notin C_k}\!\!\!\!\!
J^{ij}_{\mu\nu}(O_{i}^\mu-\langle O_i^\mu\rangle)(O_{j}^\nu-\langle O_j^\nu\rangle)]\,,\label{HS2}
\end{equation}
where the last term is the {\it residual interaction}.

For $T\rightarrow 0$, $\rho_k\rightarrow |0_k\rangle\langle 0_k|$, with
$|0_k\rangle$ the GS of $h_k$. The present scheme will then lead in this limit
to the state
\begin{equation}|0_h\rangle=\otimes_{k}|0_k\rangle\,,\end{equation}
which minimizes $\langle H\rangle\equiv \langle\Psi|H|\Psi\rangle$ among {\it
all} cluster product states $|\Psi\rangle=\otimes_k |\psi_k\rangle$. Let us
remark that an explicit ansatz for the states $|0_k\rangle$ is not required,
since they can be obtained as the GS of $h_k$, Eq.\ (\ref{hk}), in each
iteration. Nonetheless, in certain cases (see sec.\ III) the explicit form of
$|0_k\rangle$ may become apparent from the form of $h_k$ and a direct
minimization of $\langle H\rangle$ becomes feasible.

\subsection{Perturbative corrections and symmetry restoration}
While in-cluster correlations are already described by $\rho_k$ or
$|0_k\rangle$, those between clusters can in principle be estimated through
perturbative corrections. At $T=0$, it follows from Eq.\ (\ref{HS2}) that $H$
will connect $|0_h\rangle$ just with {\it two-cluster}  excitations $|n_k
n'_{k'}\rangle$, $k\neq k'$, $n n'\neq 0$, where $|n_k\rangle$ are the
eigenstates of $h_k$ ($h_k|n_k\rangle= \varepsilon_{n_k}|n_k\rangle$).
Consequently, first order (in the residual interaction) corrections will lead
to the perturbed GS
\begin{eqnarray}|0^1_H\rangle&\propto&|0_h\rangle+\sum\limits_{k<k',n,n'\geq 1}
\alpha_{kn,k'n'}|n_k n'_{k'}\rangle\,,
\label{pt}\\
\alpha_{kn,k'n'}&=&{\textstyle\sum\limits_{i\in C_k,j\in C_{k'}}
J^{ij}_{\mu\nu}\frac{\langle n_k|O_i^\mu|0_k\rangle\langle
n'_{k'}|O_{j}^\nu|0_{k'}\rangle}
{\varepsilon_{n_k}-\varepsilon_{0_k}+\varepsilon_{n'_{k'}}-
 \varepsilon_{0_{k'}}}}\,,\end{eqnarray}
which contains just two-cluster excitations.

For instance, the reduced state of cluster $k$ derived from (\ref{pt}) is
($\bar{k}$ denotes the complementary system)
\begin{equation}\rho_k={\rm Tr}_{\bar{k}}|0^1_H\rangle\langle 0^1_H|\propto
|0_k\rangle\langle
0_k|+\sum_{n,m}(\alpha\alpha^\dagger)_{kn,km}|n_k\rangle\langle m_k|\,,
 \label{rhokp}\end{equation}
which is a {\it mixed} state. Its entropy $S(\rho_k)$ represents the
entanglement of the cluster with the rest of the system.

Beyond the weak coupling limit, the actual potential of the GMF lies in the
possibility of breaking some essential symmetry of $H$, which will enable it to
describe non-perturbative coupling effects between the composite  sites. We
will be here concerned with a discrete broken symmetry, namely spin parity
symmetry $P_z$ (see next section), such that GMF will yield in some sectors a
pair of parity breaking degenerate solutions $h_\pm$, with $h_-=P_z h_+ P_z$.
We can then construct from the parity  breaking GS $|0_{h_{+}}\rangle=\otimes_k
|0_{k+}\rangle$ and $|0_{h_-}\rangle=P_z|0_{h_+}\rangle$, the definite parity
states
 \begin{equation}
 |0_{\pm}\rangle=\frac{|0_{h_{+}}\rangle\pm|0_{h_{-}}\rangle}
 {\sqrt{2[1\pm{\rm Re}(\langle 0_{h_+}| 0_{h_-}\rangle)]}}\,,\label{DP}
 \end{equation}
 which will normally be not strictly degenerate in finite systems
and which lead to a non-perturbative entanglement between composite sites:
Neglecting the complementary overlap  $\prod_{k'\neq k} \langle
0_{k'+}|0_{k'-}\rangle$, typically small, the ensuing reduced  state of the
cluster $k$ will  be the same for $|0_{\pm}\rangle$ and given by
\begin{equation}\rho_k={\rm Tr}_{\bar{k}}|0_\pm\rangle\langle 0_{\pm}|\approx
\frac{1}{2}(|0_{k+}\rangle\langle 0_{k+}|+|0_{k-}\rangle\langle
 0_{k-}|)\,,\label{ris}\end{equation}
which is a rank 2 mixed state with eigenvalues
\begin{equation}
p_{\pm}=\frac{1}{2}(1\pm|\langle 0_{k+}|0_{k-}\rangle|)\,,\label{pkpm}
\end{equation}
and non-zero entropy
$S(\rho_k)$. A parity breaking GMF is then a signature of a non-perturbative
entanglement $S(\rho_k)$ between the composite site and the rest of the system
in the exact (definite parity) GS. Similar considerations hold for
 a group $G$ of clusters, for which the reduced state will again be a
similar rank 2 mixed state with $p_{\pm}=\frac{1}{2}(1\pm\prod_{k\in G}|\langle
0_{k+}|0_{k-}\rangle|)$. For a large group, $p_{\pm}\rightarrow 1/2$ and
$S(\rho_G)\rightarrow \ln 2$. Such contribution is analogous to a ``topological''
entropy \cite{KP.06}.

\section{Application to dimerized spin systems \label{III}}
\subsection{Dimerized $XY$ spin chain}

\begin{figure}[t]
\vspace*{0cm}

\centerline{\hspace*{-0.2cm}\scalebox{.5}{\includegraphics{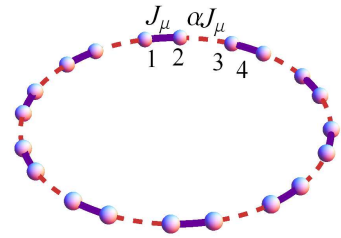}}}
\caption{(Color online) Schematic plot of the dimerized cyclic chain.}
\label{f1}
\end{figure}
We first consider a cyclic spin $1/2$ chain of $N=2n$ spins in a transverse
uniform field $B$, coupled through alternating first neighbor anisotropic $XY$
couplings \cite{Pr.75,Pr.77,FM.07,GG.09,CRM.10}, such that the system can be
viewed, at least for weak fields, as strongly coupled pairs weakly interacting
with their neighboring pairs (Fig.\ \ref{f1}). The Hamiltonian can be written
as
\begin{equation}
H=\sum_{i=1}^n [B (s^z_{2i-1}+s^z_{2i})-\sum_{\mu=x,y}J_\mu(s_{2i-1}^\mu
 s_{2i}^\mu+\alpha_\mu s_{2i}^\mu s_{2i+1}^\mu)]\,, \label{H1}\end{equation}
where $s_i^\mu$ denotes the (dimensionless) spin component at site $i$. We will focus on the
case $\alpha_x=\alpha_y=\alpha$ (common anisotropy). We can suppose, without
loss of generality, $|\alpha|\leq 1$ and, moreover,  $\alpha\geq 0$, both in a
cyclic chain ($s_{2n+1}^\mu=s_{1}^\mu$) with an even number $n$ of pairs or in
an open chain with $n$ pairs, as its sign can be changed by a rotation of angle
$\pi$ around the $z$ axis at even pairs (sites $2i-1,2i$, $i$ even)
\cite{CRM.10}. A similar rotation at all even sites changes the sign of $J_x$
and $J_y$, so that we can also assume $J_x\geq 0$, with $|J_y|\leq J_x$. We set
here $|J_y|<J_x$. Finally, we set $B\geq 0$, as its sign can be changed by a
global rotation of angle $\pi$ around the $x$ axis, which leaves the couplings
unchanged. These arguments also hold for arbitrary spin $s$.

Eq.\ (\ref{H1}) commutes with the total $S_z$ parity \begin{equation}
P_z=\exp[-i\pi (S_z+2ns)]
\label{Pz} \,,\end{equation}
where $S_z=\sum_{i=1}^{2n} s_{i}^z$. This implies  $\langle s_i^\mu\rangle=0$
for $\mu=x,y$ in any non-degenerate eigenstate. Breaking of this symmetry
($\langle s_i^\mu\rangle\neq 0$ for $\mu=x$ or $y$) is, however, essential
in MF descriptions, at least within some field intervals.

The conventional MF is based on a product state
\begin{equation}\rho_h=\otimes_{i=1}^{2n}\rho_i\,,\;\;\rho_i=\exp[-\beta h_i]/Z_{h_i}\,,
\label{mf1}\end{equation}
where, for the chosen signs of couplings, we may assume all $\rho_i$ identical
in the cyclic case, such that $\langle s_i^\mu\rangle=\langle s^{\mu}\rangle$
and $\langle H\rangle=n[2B\langle s^z\rangle-(1+\alpha)\sum_{\mu} J_\mu\langle
s^\mu\rangle^2]$, with
\begin{equation}h_i=\bm{\lambda}\cdot\bm{s}_i=  B s_i^z-(1+\alpha)
 \sum_{\mu=x,y}J_\mu\langle s^\mu\rangle s_i^\mu\,.\end{equation}
Considering now $T=0$, the GS $|0_i\rangle$ of $h_i$ will be a state with
maximum spin along $-\bm{\lambda}$, leading to $\langle
s^z\rangle=-s\cos\theta$, $\langle s^x\rangle=s\sin\theta\cos\phi$, $\langle
s^y\rangle=s\sin\theta\sin\phi$. Minimization of $\langle H\rangle$ for
$|J_y|<J_x$ leads then to $\phi=0$ ($\langle s^y\rangle=0$) and
\begin{equation}
\left\{\begin{array}{ll}\theta=0\,,&\;\;B\geq B_c^\alpha\equiv J_x(1+\alpha)s\,,\\
\cos\theta=B/B_c^\alpha\,,&\;\;B<B_c^\alpha\,,
\end{array}\right.
\label{thmf}
\end{equation}
with parity broken for $B<B_c^\alpha$, where the solution is degenerate
($\theta=\pm|\theta|$). For $s=1/2$ we then obtain
\begin{eqnarray}
\langle 0_h|H|0_h\rangle&=&-n\left\{\begin{array}{lr}B&B
\geq B_c^\alpha\\\frac{1}{2}(\frac{B^2}{B_c^\alpha}+B_c^\alpha)&B<B_c^\alpha
\end{array}\right.\,,
\label{EMF}
\end{eqnarray}
where $|0_h\rangle=\otimes_{i=1}^{2n}|0_i\rangle$ with (Fig.\ \ref{f2})
\begin{eqnarray}|0_i\rangle&=&
{\textstyle\cos\frac{\theta}{2}|\downarrow\rangle+\sin\frac{\theta}{2}
|\uparrow\rangle\,.}\label{0mf}\end{eqnarray}
This simple approach ignores the dimerized structure of the chain (it is the
same as that for a chain with uniform coupling $J_x(1+\alpha)/2$), and is also
blind to the weaker $J_y$ coupling. Yet, it is remarkable that if $J_y\geq 0$,
$|0_h\rangle$  does become an {\it exact} GS at the {\it separability field}
\cite{GG.09,CRM.10,GAI.09,RCM.08,KTM.82}
\begin{equation}B_s^\alpha\equiv
\sqrt{J_y J_x}(1+\alpha)s=\sqrt{J_y/J_x}\,B_c^\alpha\,,\label{BS}\end{equation}
where $\cos\theta=\sqrt{J_y/J_x}$. At this field  the system  exhibits a {\it
degenerate} GS, with the GS subspace spanned by the pair of degenerate MF
product states \cite{CRM.10,RCM.08}. No traces of dimerization are left at this
point in the exact GS.

\subsection{Pair Mean Field Approximation}
In order to improve the conventional MF picture for $B\neq B_s^\alpha$, we now
examine a generalized MF approach based on {\it independent spin pairs}, such
that
\begin{equation}\rho_h=\otimes_{i=1}^n\rho_i^p\,,\;\;\rho_i^p=\exp[-\beta h_i^p]/Z_{h_i^p}\,,
\label{mfp}\end{equation}
with $\rho_i^p$ a pair state. Eq.\ (\ref{mfp}) is exact in the fully dimerized
limit $\alpha\rightarrow 0$, and can then be expected to provide a good
approximation at least for small  $\alpha$. For the chosen signs of couplings,
we may again assume all $\rho_{i}^p$ identical in the cyclic case, with
$\langle s_i^\mu\rangle=\langle s^\mu\rangle$, implying
\begin{equation}\langle H\rangle=
n[ 2B \langle s^z\rangle-\sum_{\mu=x,y} J_\mu
(\langle s_1^\mu s_2^\mu\rangle+\alpha\langle s^\mu\rangle^2)]
\,, \label{Hp}\end{equation}
and
\begin{equation}h^p_i=B (s_{2i-1}^z+s_{2i}^z)-\sum_{\mu=x,y} J_\mu
[s_{2i-1}^\mu s_{2i}^\mu+\alpha\langle s^\mu\rangle(s_{2i-1}^\mu+
 s_{2i}^\mu)]\,.\label{hp}\end{equation}
For $|J_y|<J_x$, minimization of $\langle H\rangle$ leads again to $\langle s^y\rangle=0$.

In the case of arbitrary spin and temperature, one should start from an initial
seed for $\langle s^x\rangle$, diagonalize $h^p_i$ and then recalculate $\langle
s^x\rangle$ until convergence is reached. Considering now $T=0$ and $s=1/2$, it
is apparent from (\ref{hp}) that the GS of $h_i^p$ will  be of the form
\begin{eqnarray}
\!\!\!\!\!\!\!\!\!\!|0_{i}^p\rangle&=&{\textstyle\cos\frac{\theta}{2}(\cos\frac{\phi}{2}|\downarrow\downarrow\rangle+
\sin\frac{\phi}{2}|\uparrow\uparrow\rangle)+\sin\frac{\theta}{2}
\frac{|\uparrow\downarrow\rangle+|\downarrow\uparrow\rangle}{\sqrt{2}}}\,,\label{pst}
\end{eqnarray}
which is just the most general symmetric pair state real in the standard basis.
Eq.\ (\ref{Hp}) becomes
\begin{eqnarray}
\langle 0_h^p|H|0_h^p\rangle
&=&{\textstyle-n[(B\cos\phi+J_-\sin\phi)\cos^2\frac{\theta}{2}+J_+\sin^2\frac{\theta}{2}
}\nonumber\\
&&+{\textstyle\frac{1}{8}\alpha J_x\sin^2\theta(1+\sin\phi)]}\,,\label{Ep}
\end{eqnarray}
where $|0_{h}^p\rangle=\otimes_{i=1}^n |0_i^p\rangle$ and $J_{\pm}=\frac{J_x\pm J_y}{4}\geq 0$.
Minimization of $\langle H\rangle$ with respect to $\theta,\phi$ can then be directly done, leading to
\begin{subequations}\label{ss}
\begin{eqnarray}
\;\;\theta=0,\;\tan\phi=\frac{J_-}{B}\,,&\;\;\; &B\geq B_{c2}^\alpha\,,\label{a}\\
\!\!\!\!\!\!\!\!\!{\left\{\begin{array}{l}\cos \theta={2\frac{B\cos\phi+J_-\sin\phi-J_+}{\alpha J_x(1+\sin\phi)}}\\
\tan\phi=\frac{J_- +\alpha J_x(1-\cos\theta)/4}{B}\end{array}\right.
}\,,&\;\;\;&
B_{c1}^\alpha< B< B_{c2}^\alpha\,,\label{b}\\
\;\;\theta=\pi\;(\phi\;{\rm arbitrary})\,,&\;\;\;&B\leq B_{c1}^\alpha\,\label{c}
\end{eqnarray}
\end{subequations}
where the critical fields are given by
\begin{eqnarray}
B_{c1}^\alpha&=&{\textstyle\frac{1}{2}\sqrt{J_x(J_y-2\alpha J_x)}}\,,\label{Bc1}\\
B_{c2}^\alpha&=&{\textstyle\frac{1}{2}\sqrt{(J_++\frac{\alpha}{2}J_x+
\sqrt{(J_++\frac{\alpha}{2}J_x)^2+2\alpha J_x J_-})^2-4J_-^2}}\,,\nonumber\\&&
 \label{Bc2}\end{eqnarray}
as obtained from (\ref{b}) for $\theta\rightarrow 0$ and
$\theta\rightarrow\pi$. The  solution of system (\ref{b}) for $\theta$ and
$\phi$ can in fact be determined analytically (it leads to a quartic equation
for $\cos\phi$).

\begin{figure}[t]
\vspace*{0cm}

\centerline{\hspace*{-0.2cm}\scalebox{.8}{\includegraphics{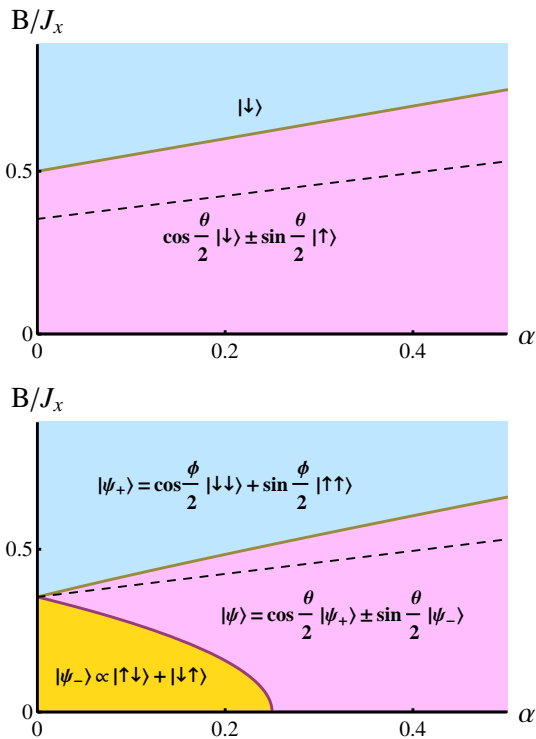}}}
\caption{(Color online) Phase diagram of the dimerized spin $1/2$ chain
according to the conventional (top panel) and  pair (bottom panel) mean field
approaches, for $J_x>0$ and $J_y=J_x/2$. The corresponding states for the unit
cell are indicated. While in the conventional MF the $S_z$ parity breaking
phase arises below a critical field $B_c^\alpha$, in the pair MF it occurs
within a field window $B_{c1}^\alpha<B<B_{c2}^\alpha$ if $\alpha<\alpha_c$
(Eq.\ (\ref{ac})). For $B<B_{c1}^\alpha$ a dimerized state with maximally
entangled pairs is preferred. The dashed lines denote the factorizing field
$B_s^\alpha$ where both MF approaches coincide and are exact.} \label{f2}
\end{figure}

In contrast with the standard MF, it is first seen that a parity breaking
solution ($\theta\in(0,\pi/2)$) will now arise just within a {\it field window}
$B_{c1}^\alpha<B<B_{c2}^\alpha$ if $\alpha$ is sufficiently small and $J_y>0$,
as depicted in Fig.\ \ref{f2} (bottom panel). For $B<B^\alpha_{c1}$, the pair
MF leads to a {\it fully dimerized phase},  where the strongly coupled pairs
are in a $P_z=-1$ {\it Bell state}
$\frac{|\uparrow\downarrow\rangle+|\downarrow\uparrow\rangle}{\sqrt{2}}$ and
hence maximally entangled. On the other hand, for $B>B^\alpha_{c2}$ the
approach leads to an entangled $P_z=1$ pair state $\cos\frac{\phi}{2}
|\!\!\downarrow\downarrow\rangle
+\sin\frac{\phi}{2}|\!\!\uparrow\uparrow\rangle$, which is only partially
aligned. The intermediate parity breaking phase (\ref{b}) is then a transition
region between the previous opposite parity phases, in which the pair is in a
combination of the previous states. In this region the pair MF GS is two-fold
degenerate ($\theta=\pm|\theta|$). It is verified that the actual exact GS
obtained from the Jordan-Wigner fermionization also exhibits two transitions
for increasing positive fields \cite{Pr.75} if $\alpha$ is sufficiently small,
becoming in a finite chain nearly two-fold degenerate in the intermediate
sector\cite{GG.09,CRM.10} and leading as well to almost maximally entangled
pairs for low fields (see appendix and next section).

In the parity preserving phases, the pair MF GS energy obtained from (\ref{Ep})
is just
\begin{equation}
 \langle 0^p_h|H|0^p_h\rangle=-n\left\{\begin{array}{lr}\sqrt{B^2+J_-^2}
 \,,&B\geq B_{c2}^\alpha\\J_+\,,&B\leq B_{c1}^\alpha\end{array}\right.\,,
\label{EPMF}
\end{equation}
which is, of course, lower than the conventional MF energy (\ref{EMF}) in
these intervals.

The factorizing field (\ref{BS}) lies within the parity breaking phase
$\forall$ $\alpha>0$: $B_{c1}^\alpha<B_{s}^\alpha<B_{c2}^\alpha$. It is
verified that at $B=B_s^\alpha$,  Eq.\ (\ref{b}) leads  to $\cos\theta=J_y/J_x$
and $\tan\phi=2J_-/\sqrt{J_xJ_y}$, implying
\begin{equation}\tan^2\theta/2=\sin\phi\,,\label{eqbs}\end{equation}
which is precisely  the condition ensuring that the pair state
(\ref{pst}) reduces to a product of single spin states.

On the other hand, for $\alpha\rightarrow 0$ (where the pair MF becomes exact),
$B_{c1}^\alpha$ and $B_{c2}^\alpha$ merge (Fig.\ \ref{f2}), approaching both
the $\alpha=0$ factorizing field $B_s^0=\sqrt{J_y J_x}/2$ ($B_{c
1,2}^\alpha\approx B_s^0(1\mp\alpha\frac{J_y}{J_x})$ for small $\alpha$): The
exact GS of an isolated pair undergoes, for $J_y>0$, a sharp parity transition
at $B=B_s^0$, from the Bell state
$\frac{|\uparrow\downarrow\rangle+|\downarrow\uparrow\rangle}{\sqrt{2}}$ for
$B<B_s^0$, with energy $-J_+$ (Eq.\ (\ref{EPMF})) to the state
$\cos\frac{\phi}{2} |\!\!\downarrow\downarrow\rangle
+\sin\frac{\phi}{2}|\!\!\uparrow\uparrow\rangle$ for $B>B_s^0$, with energy
$-\sqrt{B^2+J_-^2}$. At $B=B_s^0$ these states become degenerate and coincide
with the definite parity combinations (\ref{DP}) of the MF product states
$\otimes_{i=1}^{2}|0_i\rangle$.

It is also seen from Eq.\ (\ref{Bc1}) that $B_{c1}^\alpha$ vanishes for
\begin{equation}
\alpha=\alpha_c\equiv\frac{J_y}{2J_x}\,.\label{ac}
\end{equation}
If $\alpha>\alpha_c$ (or $J_y<0$) parity is broken for all $B\leq
B_{c2}^\alpha$, as in the standard MF. Nonetheless, important differences with
the latter persist: $B_{c2}^\alpha$ remains lower than the MF critical field
$B_c^\alpha$, even for $\alpha=1$, and strongly coupled pairs remain entangled
even for strong fields $B>B^\alpha_{c}$: Full alignment occurs only for
$B\rightarrow\infty$, with $\phi\approx J_-/B$ for $B\gg J_-$. The pair MF
depends also on $J_y$, which affects the critical fields and the values of
$\theta,\phi$.

If $J_y<0$ (with $|J_y|\leq J_x$), $B_{c2}^\alpha$ also
vanishes at $\alpha=-\frac{J_y}{2J_x}\geq 0$, entailing {\it no parity breaking
phase} in the pair MF if $\alpha\leq-\frac{J_y}{2J_x}$. This is in qualitative
agreement with the exact result (see appendix), but differs from the standard
MF, where parity breaking still occurs $\forall$ $\alpha$.

If $J_x>0$ but $\alpha<0$, the pair MF state can be obtained by rotation of
angle $\pi$ around the $z$ axis at {\it even} pairs of the $\alpha>0$ pair
state, which implies (ignoring in what follows overall phases) an {\it
alternating} angle $\theta$ in (\ref{pst}) ($\theta_i=(-1)^i\theta$) in the
parity breaking phase. If $\alpha>0$ but $J_x<0$ (with $|J_y|<|J_x|$), such
rotation should be applied to all even sites, entailing
$|\!\!\uparrow\downarrow\rangle+|\!\!\downarrow\uparrow\rangle\rightarrow
 |\!\!\uparrow\downarrow\rangle-|\!\!\downarrow\uparrow\rangle$ and
 $\cos\frac{\phi}{2}|\!\!\downarrow\downarrow\rangle+\sin\frac{\phi}{2}
 |\!\!\uparrow\uparrow\rangle\rightarrow
 \cos\frac{\phi}{2}|\!\!\downarrow\downarrow\rangle-\sin\frac{\phi}{2}
|\!\!\uparrow\uparrow\rangle$
 in (\ref{pst}).

 \subsection{Entanglement predictions and comparison with exact results}

\begin{figure}
\vspace*{0cm}

\centerline{\hspace*{-0.2cm}\scalebox{.78}{\includegraphics{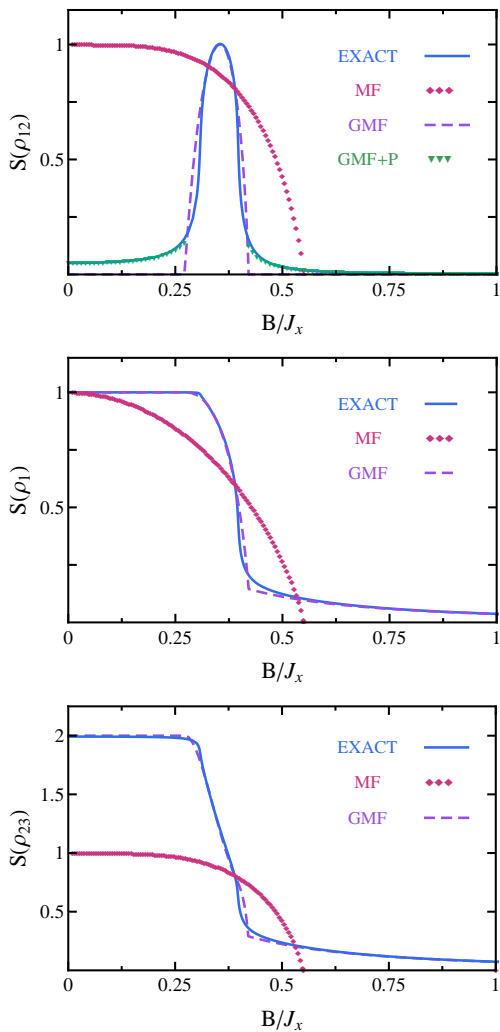}}}
\caption{(Color online) Exact and approximate results for the GS entanglement
entropy of a strongly coupled spin pair (top), a single spin (center) and a
weakly coupled neighboring pair (bottom), with the rest of the chain, for
$\alpha=0.1$ and $J_y/J_x=1/2$, as a function of the (scaled) magnetic field.
MF denotes the conventional single spin MF treatment (\ref{mf1})--(\ref{0mf}),
while GMF the pair MF approach (\ref{mfp})--(\ref{pst}), both with symmetry
restoration (Eq.\ (\ref{ris})), and GMF+P the perturbed pair MF approach
(\ref{pt})--(\ref{rhokp}). } \label{f3}
\end{figure}

\begin{figure}[t]
\centerline{\hspace*{-0.2cm}\scalebox{.8}{\includegraphics{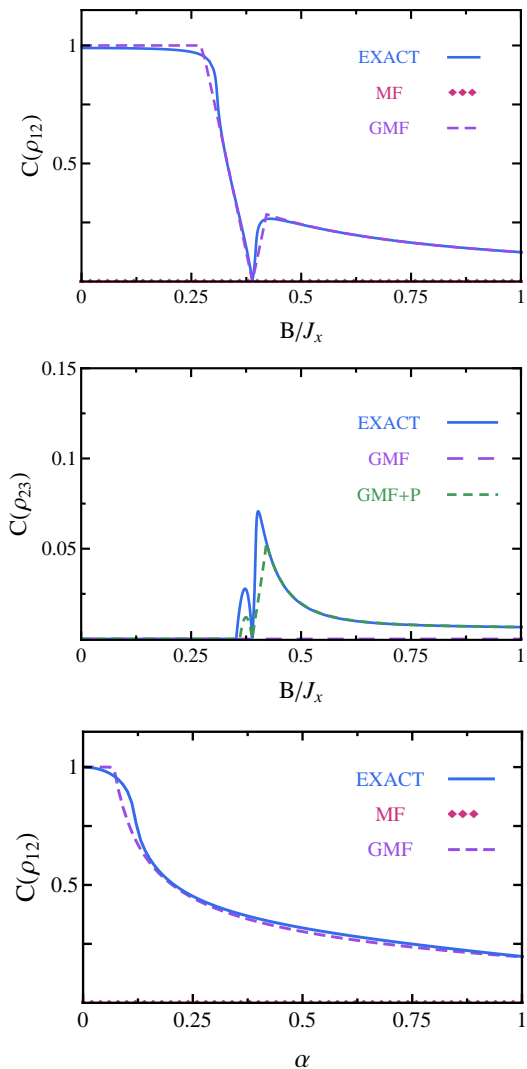}}}
\caption{(Color online) The concurrence of a strongly (top) and weakly (center)
coupled pair of neighboring spins, as a function of the scaled magnetic field
for the chain of Fig.\ \ref{f5}, according to exact and approximate results.
The bottom panel depicts the concurrence of a strongly coupled pair as a
function of the weak coupling parameter $\alpha$, at fixed field $B=0.3 J_x$.
The standard  MF result vanishes in all panels. } \label{f4}
\end{figure}

We first show in Figs.\ \ref{f3}-\ref{f4} typical GS results for different
entanglement observables related with spin pairs and single spins in a finite chain
with $n=50$  pairs, according to conventional and pair MF as well as
exact results (see Appendix). The latter correspond to the exact GS of the
finite chain (having then a definite $S_z$ parity).

 For a pair of strongly coupled neighboring spins (1-2 in Fig.\ \ref{f1}), the
pair MF approach (\ref{mfp}) leads, after the symmetry restoration
(\ref{DP})--(\ref{ris}), to the reduced state
\begin{eqnarray}
\rho_{12}^{_{\rm GMF}}=
\left( \begin{array}{cccc}\cos^2\frac{\theta}{2}\sin^2\frac{\phi}{2} &0&0&
\frac{1}{2} \cos^2\frac{\theta}{2}\sin\phi\\
0&\frac{1}{2} \sin^2\frac{\theta}{2}&\frac{1}{2}\sin^2\frac{\theta}{2}&0\\
0&\frac{1}{2}\sin^2\frac{\theta}{2}&\frac{1}{2}\sin^2\frac{\theta}{2}&0\\
\frac{1}{2} \cos^2\frac{\theta}{2}\sin\phi &0&0 &\cos^2\frac{\theta}{2}\cos^2
\frac{\phi}{2}\end{array}\right)
\,,
\nonumber\\\label{rhomfg12}
\end{eqnarray}
(expressed in the std.\ basis) after neglecting the overlap $|\langle
0_i^p(\theta)|0_i^p(-\theta)\rangle|^{n-1}$ in the parity breaking
phase. In this phase it is a rank $2$ mixed state (and is pure otherwise), with
eigenvalues $(\sin^2\frac{\theta}{2},\cos^2\frac{\theta}{2})$ (Eq.\
(\ref{pkpm}). It then leads within this phase to a {\it non-zero entanglement
entropy} $E_{12}=S(\rho_{12})$ between the pair and the rest of the chain. As
seen in the top panel of Fig.\ \ref{f3}, this is in agreement with the exact
result, which also exhibits a pronounced peak in this interval (we use
$S(\rho)=-{\rm Tr}\rho\log_2\rho$ in all panels). Parity breaking in the pair
MF is then a signature of a {\it non-negligible entanglement} between this pair
and the rest of the chain. The exact result presents as well small nonzero
tails outside the parity breaking interval, which can be correctly predicted by
the {\it perturbed} pair MF reduced state  (\ref{rhokp}).

Note that the entropy $S(\rho_{12})$ does not vanish as $B$ approaches the
factorizing field $B_s^\alpha$ ($\approx 0.39 J_x$ in Fig.\ \ref{f3}), since
the exact GS remains with a definite parity (and hence entangled) in its
immediate vicinity.  In fact, for $B\rightarrow B_s^\alpha$  the result
obtained from (\ref{rhomfg12}) becomes {\it exact} (except for the small
neglected overlap), as the parity restored pair MF GS is exact in this limit.
Actually, as stated before, at $B=B_s^\alpha$ the exact GS is degenerate, so
that GS entanglement will depend at this point on the choice of GS. The result
obtained from (\ref{rhomfg12}) corresponds to the definite parity GS's
(\ref{DP}), which are the {\it actual side limits} \cite{RCM.08} of the exact
GS  for $B\rightarrow {B_s^\alpha}^\pm$.

The single spin state derived from (\ref{rhomfg12}) is just
\begin{equation}
\rho_1^{_{\rm
GMF}}=\left(\begin{array}{cc}p_+&0\\0&p_-\end{array}\right)\,,\;\;
p_{\pm}={\textstyle\frac{1}{2}(1\mp\cos^2\frac{\theta}{2}\,\cos\phi)}\,,
 \label{rho1}\end{equation}
which is of the form $\frac{1}{2}(\rho_1^++\rho_1^-)$ in the parity breaking
phase,  with $\rho_1^{\pm}$ the single spin reduced states derived from the
pair state (\ref{pst}) before parity restoration. Its entropy, quantifying its
entanglement with the rest of the chain, is non-zero for all fields and seen to
be almost coincident with the exact result (center panel). It is obviously
maximum in the dimerized phase $B<B^\alpha_{c1}$, but decreases rapidly in the
parity breaking phase (when the pair becomes entangled with the rest of the
chain) and slowly in the partially aligned phase $B>B^\alpha_{c2}$ (where
$p_+\approx \phi^2/4=J_-^2/(4B^2)$). The result derived from (\ref{rho1}) is
again fully exact for $B\rightarrow B_s^\alpha$.

The entanglement entropy $S(\rho_{23})$ of a weakly coupled pair with the rest
of the chain can again be correctly described by the pair MF approach, as seen
in the bottom panel. Note that $\rho_{23}^{_{\rm
GMF}}=\frac{1}{2}(\rho_1^+\otimes \rho_1^++\rho_1^-\otimes\rho_1^-)$, so that
in the parity preserving phases ($\rho_1^{+}=\rho_1^-$), $S(\rho_{23}^{_{\rm
GMF}})$ is just twice the single spin entropy $S(\rho_1^{_{\rm GMF}})$. This
relation no longer holds, however, in the parity breaking phase.

In contrast, it is verified in all panels that the conventional MF (\ref{mf1})
does not lead to a proper picture of any of these  measures, even
after symmetry restoration. The ensuing reduced pair state is the same for any
pair,
 \begin{eqnarray}
\;\;\rho_{12}^{_{\rm MF}}&=&
\left( \begin{array}{cccc}\sin^4\frac{\theta}{2} &0&0&\frac{1}{4}\sin^2\theta\\
0&\frac{1}{4}\sin^2\theta&\frac{1}{4}\sin^2\theta&0\\
0&\frac{1}{4}\sin^2\theta&\frac{1}{4}\sin^2\theta&0\\
\frac{1}{4}\sin^2\theta &0&0 &\cos^4\frac{\theta}{2}\end{array}\right)\,,\label{rho12}
\end{eqnarray}
which is a rank $2$ state for $\theta\in(0,\pi)$ with eigenvalues $\frac{1\pm
\cos^2\theta}{2}$ ($\theta$ is here the MF angle (\ref{thmf})). Its entropy
does not reflect the exact entanglement of the strongly nor the weakly coupled
pair. The associated single spin reduced state is of the form (\ref{rho1}) but
with $p_{\pm}=(1\mp\cos\theta)/2$, and cannot correctly reproduce either its
entanglement with the rest of the chain (center panel in Fig.\ \ref{f3}). It
is seen, however, that there is one point where the conventional MF result is
exact for all three quantities (i.e., where the MF curve crosses the exact
curve), which is the factorizing field $B_s^\alpha$. Here the reduced states
(\ref{rho12}) and (\ref{rhomfg12}) become identical and, moreover, exact.

Fig.\ \ref{f4} depicts the {\it concurrence} \cite{Wo.97}, a measure of the
entanglement {\it between} the spins of pair, for both strongly (1-2) and
weakly (2-3) coupled pairs. In the first case, the pair MF state
(\ref{rhomfg12}) leads to  the concurrence
\begin{equation}C(\rho^{_{\rm GMF}}_{12})=|\cos^2{\textstyle\frac{\theta}{2}}\,
   (1+\sin\phi)-1|\,.\label{C12}\end{equation}
which is parallel (as that in a state
$|\!\!\uparrow\uparrow\rangle+|\!\!\downarrow\downarrow\rangle$) if the term
within the bars is positive, i.e., $B>B_s^\alpha$, and antiparallel (as that in
$|\!\!\downarrow\uparrow\rangle+|\!\!\uparrow\downarrow\rangle\!$) if this term
is negative,  i.e., $B<B_s^\alpha$, {\it vanishing} at the factorizing field
$B_s^\alpha$ (see below). As seen in the top panel, the pair MF result shows
again a very good agreement with the exact result for all fields, correctly
predicting a maximally entangled pair for low fields $B<B_{c1}^\alpha$. Note
that for $B<B_{c1}^\alpha$  and $B>B_{c2}^\alpha$, the state (\ref{rhomfg12})
is pure, implying that the pair MF concurrence is just a function of
$S(\rho_{1}^{_{\rm GMF}})$, and given by
\begin{equation}
C^{_{\rm GMF}}_{12}=\left\{\begin{array}{ccl}1&,&B<B_{c1}^\alpha\\
\frac{J_-}{\sqrt{B^2+J_-^2}}&,&B>B_{c2}^\alpha\label{C122}\end{array}
\right.\,,
\end{equation}
decreasing as $J_-/B$ for strong fields $B\gg J_-$. However, in the parity
breaking phase the state (\ref{rhomfg12}) is mixed and the concurrence
(\ref{C12}) is no longer a function of $S(\rho_1^{_{\rm GMF}})$. In fact, and
as opposed to the previous entropies, it {\it vanishes} at the factorizing
field $B_s^\alpha$, as can be verified from Eqs.\ (\ref{eqbs}), (\ref{C12}),
since the state (\ref{rhomfg12}) becomes {\it separable} (a convex combination
of product states \cite{WW.89}) at this point. Here the single spin ceases to
be entangled with its partner (except for tiny overlap corrections) even though
it remains entangled with the rest of the chain ($S(\rho_1^{_{\rm GMF}})\neq
0$), indicating again that no traces of dimerization remain.

We also mention that the fidelity \cite{NC.00} of the state (\ref{rhomfg12})
with the exact $\rho_{12}$, $F={\rm Tr}\sqrt{\sqrt{\rho_{12}}\rho^{_{\rm
GMF}}_{12}\sqrt{\rho_{12}}}$, is very high ($\agt 0.99$ for $\alpha=0.1$ in all
phases). In contrast, the conventional MF state (\ref{rho12}) has a low
fidelity, especially for  $B<B_{c2}^\alpha$, and leads to a {\it zero}
concurrence $\forall$ $B$, since it is a separable state even after parity
restoration ($\rho_{12}^{_{\rm MF}}=\frac{1}{2}(\tilde{\rho}_1^+\otimes
\tilde{\rho}_1^++\tilde{\rho}_1^-\otimes \tilde{\rho}_1^-)$, with
$\tilde{\rho}_1^\pm$ the MF single spin state before parity restoration).

The concurrence of a weakly coupled neighboring pair is plotted in the central
panel of Fig.\ \ref{f4}. This quantity cannot be reproduced by the standard nor
the pair MF, since even after parity restoration they lead to a separable state
$\rho_{23}$. However,  it can be correctly described by the reduced state
$\rho_{23}^{_{\rm GMF+P}}$ derived from the perturbed pair MF state (\ref{pt}).
This concurrence is small and starts to be non-zero just before the factorizing
field $B^\alpha_s$, having peaks at both sides of $B^\alpha_s$. We should
actually recall that at the immediate vicinity of $B^\alpha_s$ (i.e.,
$B\rightarrow B_s^{\alpha\pm}$), the concurrence between {\it any} two spins
acquires in a finite chain a common tiny yet non-zero value in the definite
parity GS, which can be exactly predicted by both the conventional or pair MF
after parity restoration  if the overlap $|\langle \psi_{\theta\phi}|
\psi_{-\theta\phi}\rangle|^{n-1}$ is conserved \cite{CRM.10,RCM.08}.

While the general accuracy of the pair MF approach will decrease as $\alpha$
increases, it will still improve the conventional MF results, even in the
uniformly coupled case $\alpha=1$. In the bottom panel of Fig.\ \ref{f4} we
depict the pair MF concurrence of a strongly coupled pair for increasing
$\alpha$ at a fixed field, which is seen to remain accurate for all $\alpha\leq
1$. The conventional MF result vanishes $\forall \alpha$.

\begin{figure}[t]
\centerline{\hspace*{-0.2cm}\scalebox{.8}{\includegraphics{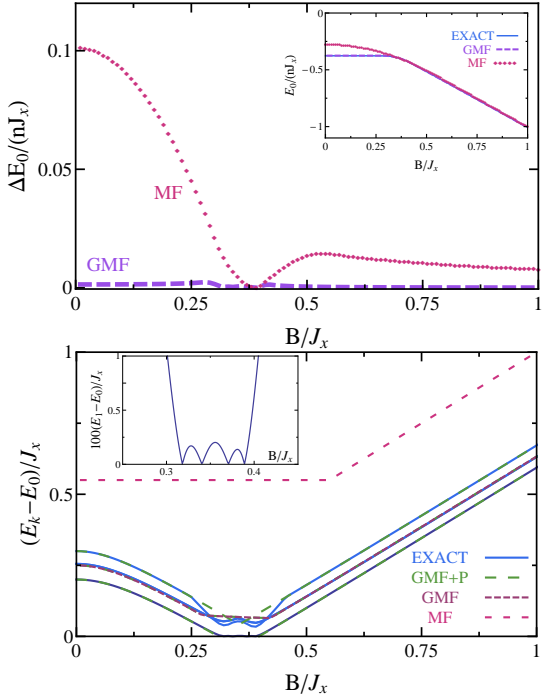}}}
\caption{(Color online) Top: The difference $\Delta E_{\rm 0}/n=(E^{\rm
app}_{\rm 0}-E^{\rm ex}_{\rm 0})/n$ between the approximate and exact GS
energies per pair, according to conventional and pair MF approaches, for the
chain of Fig.\ \ref{f4}. The inset depicts for reference the corresponding
intensive GS energies. Bottom: The first excitation energies of a small chain
with $8$ spins with the same parameters, according to pair MF and exact
results. The inset depicts a blow up of the exact first excitation energy in
the parity breaking region.} \label{f5}
\end{figure}

\subsection{Energy predictions}
We plot in Fig.\ \ref{f5} some basic energy level predictions, in
order to provide a general view of the pair MF approach. As seen in the top  panel,
the pair MF GS energy significantly improves the conventional MF result, especially
for $B<B^\alpha_s$. In the bottom panel,  we depict for clarity the first four
excitation energies in a small chain of $8$ spins ($n=4$).  According to the
pair MF approach, the lowest levels are single pair excitations, of energies
$E^{0}_m=\varepsilon_{m}-\varepsilon_{0}$ (using the notation of Eq.\
(\ref{pt})),  which in the present case will be independent of the site and
hence $n-$fold degenerate. It is verified that for small $\alpha$, this is
approximately the case. Moreover, the splitting of these levels due to the
residual interaction can be correctly described by simple first order
perturbative treatment. In the present cyclic case with a uniform pair MF, this
leads to the perturbed pair excitation energies
\begin{equation}
E^{1k}_m=\varepsilon_m-\varepsilon_0 -2\alpha\sum_{\mu=x,y}J_\mu\langle
0|s_{1}^\mu|m\rangle\langle m|s_2^\mu|0\rangle\cos{\textstyle\frac{2\pi
 k}{n}}\,, \label{Epk}\end{equation}
where $\varepsilon_m$ are the eigenvalues of the single pair Hamiltonian
(\ref{hp}) ($h^p|m\rangle=\varepsilon_m|m\rangle$), with $\varepsilon_0$ its GS
energy, and $k=1,\ldots,n$. These energies are those of  the (discrete) Fourier
transformed states $|\tilde{m}_k\rangle=\frac{1}{\sqrt{n}}\sum_{j=1}^n e^{i2\pi
kj/n}|m_j\rangle$, where $|m_j\rangle$ denotes the state with pair $j$ at
excited level $m$. As seen in the bottom panel, the result obtained from
(\ref{Epk}) is practically exact in the parity preserving phases, where the
energies $\varepsilon_m$ are  $\pm J_+$  and $\pm\sqrt{B^2+J_-^2}$, and the
lowest energies  (\ref{Epk})  become
 \begin{equation}E^{1k}_1=
 {\textstyle \pm(J_+-\sqrt{B^2+J_-^2})-\alpha (J_++\frac{J_-^2}{\sqrt{B^2+J_-^2}})
 \cos\frac{2\pi k}{n}}\,,
 \end{equation}
with $+$ for $B<B_{c1}^\alpha$ and $-$ for $B>B_{c2}^\alpha$. For $n=4$,
$E_1^{11}=E_1^{13}=E_1^0$, so that just three levels are seen. In contrast, the
conventional MF leads to a single spin excitation energy $E_1^{_{\rm MF}}=B$
for $B>B_c^\alpha$ and $J_x(1+\alpha)/2$ if $B<B_c^\alpha$, which lies well
above the previous levels.

The parity breaking phase of the pair MF approach is seen (bottom panel)
to coincide approximately with the region where the exact GS of the finite
chain becomes nearly degenerate \cite{Pr.75,Pr.77,GG.09,CRM.10}. The exact
lowest energy levels of each parity sector become very close in this interval,
actually crossing at $n$ fields (as seen in the inset), with the last crossing
taking place exactly at the factorizing field $B_s^\alpha$. This interval is
enclosed by the fields $B_{c1}^{\rm ex}$ and  $B_{c2}^{\rm ex}$ where the
lowest quasiparticle energy of the Jordan-Wigner fermionized Hamiltonian
vanishes (see appendix).

\subsection{Longer range couplings and lattices}
\begin{figure}[t]
\vspace*{0cm}

\centerline{\hspace*{-0.2cm}\scalebox{.7}{\includegraphics{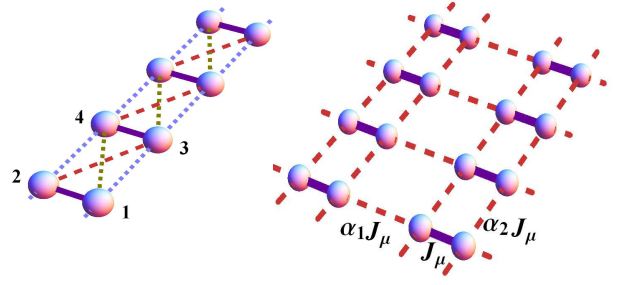}}}
\caption{(Color online) The dimerized systems corresponding
to Hamiltonians (\ref{H2}) (left) and (\ref{H3}) (right).} \label{f6}
\end{figure}

The pair MF approach remains directly applicable to more complex situations
where exact analytic results are no longer available. For instance, if adjacent
dimers in Fig.\ \ref{f1} are further connected by second and third neighbor
couplings $-\alpha_2J_\mu s^\mu_{i}s^\mu_{i+2}$ (for spins like 1-3 and 2-4)
and $-\alpha_3 J_\mu s^\mu_{2i-1}s^\mu_{2i+2}$ (for spins like 1-4), such that
\begin{eqnarray}
H&=&\sum_{i=1}^n \{B (s^z_{2i-1}+s^z_{2i})-\sum_{\mu=x,y}J_\mu
[s^\mu_{2i-1}s_{2i}^\mu\nonumber\\
&&+\sum_{j=1,2}(\alpha_{j}s_{2i}^\mu s^\mu_{2i+j}
+\alpha_{j+1}s^\mu_{2i-1}s^\mu_{2i+j})]\}\,,
  \label{H2}\end{eqnarray}
the Jordan-Wigner transformation will no longer lead  to a quadratic (and hence
analytically solvable) fermionic Hamiltonian. However, it is seen from Eqs.\
(\ref{6})--(\ref{hk}) that the previous MF and pair MF expressions and phase
diagram (Fig.\ \ref{f2}) remain valid with the replacement
\begin{equation}\alpha=\alpha_{1}+2\alpha_2+\alpha_3\,,\label{rep}\end{equation}
provided $\alpha_{2}$ and $\alpha_3$ are also positive (as $\alpha_1$) or
sufficiently small. The system of Eq.\ (\ref{H2}) is equivalent to a
ladder-type dimer chain (Fig.\ \ref{f6}, left). A uniform factorizing field
will still exist in this system for common anisotropy \cite{CRM.10,RCM.08}
($\alpha_j^\mu=\alpha_j$ $\forall$ $j$, as considered in (\ref{H2})), which
will be again given by Eq.\ (\ref{BS}) with the previous value of $\alpha$.
Similar considerations hold for longer range $XY$ couplings.

The phase diagram of Fig.\ \ref{f2} also applies, at the pair MF level, to
ferromagnetic-type $XY$ dimer lattices  like that of Fig.\ \ref{f6}, right,
described by the Hamiltonian
\begin{eqnarray} H&=&\sum_{i,j} \{B
(s^z_{2i-1,j}+s^z_{2i,j})-\sum_{\mu=x,y} J_\mu [s_{2i-1,j}^\mu
s_{2i,j}^\mu\nonumber\\&&+\alpha_1 s_{2i,j}^\mu s_{2i+1,j}^\mu +
 \alpha_2(s^\mu_{2i-1,j}s^\mu_{2i-1,j+1}+s^\mu_{2i,j} s^\mu_{2i,j+1})]\}\,,\nonumber\\
  &&\label{H3}\end{eqnarray}
where we assumed first neighbor couplings. For $\alpha_1>0$, $\alpha_2>0$, we
should just replace
\begin{equation} \alpha=\alpha_1+2\alpha_2\,,\label{alxy}\end{equation}
in the MF and pair MF approaches. Similar considerations hold for 3D lattices
or longer range couplings

\begin{figure}[t]
\centerline{\hspace*{-0.2cm}\scalebox{.7}{\includegraphics{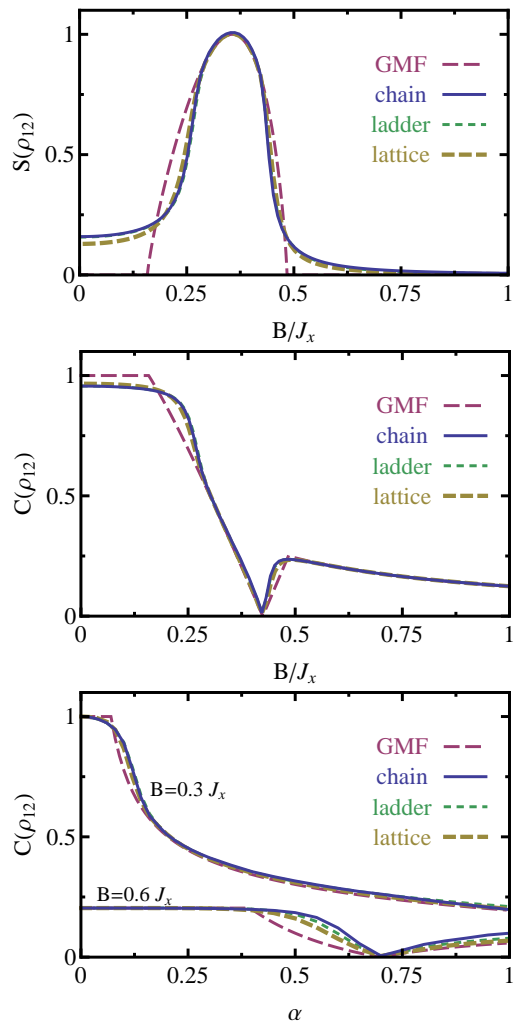}}}
\caption{(Color online) Results for the spin ladder and lattice of Fig.\
\ref{f6} (Eqs.\ (\ref{H2}), (\ref{H3})). The entanglement of strongly coupled
pairs with the rest of the system  $S(\rho_{12})$ (top), and their concurrence
$C(\rho_{12})$ (center), are plotted for increasing fields for a common value
$\alpha=0.2$ (Eqs.\ (\ref{rep})--(\ref{alxy})). Results for both systems are
very close and almost coincident with those for the cyclic chain of Fig.\
\ref{f1},  also depicted, in agreement with the common pair MF prediction
(GMF). The bottom panel depicts the concurrence for increasing values of the
total coupling parameter $\alpha$, for two fixed values of the field.}
\label{f7}
\end{figure}

Fig.\ \ref{f7} depicts illustrative results for a finite spin ladder and
lattice with cyclic conditions ($n+1=n$ in (\ref{H2}), $n_i+1=n_i$ for $i=1,2$
in (\ref{H3})). We have computed the exact results by exact diagonalization for
a total of  $2n=16$ spins ($2\times 8$ ladder, $4\times 4$ lattice). We have
set a {\it fixed} value $\alpha=0.2$ in Eqs.\ (\ref{rep}) and (\ref{alxy}),
with $\alpha_{1}=\alpha_{2}=\alpha_{3}$ in (\ref{rep}) and $\alpha_1=\alpha_2$
in (\ref{alxy}). For comparison, results for the chain of Eq.\ (\ref{H1}) with
the same $\alpha$ and spin number are also depicted.

It is verified that for a common total $\alpha$, these systems do exhibit
almost coincident values of the entanglement of a strongly coupled pair with
the rest of the system, and of its concurrence, confirming the pair MF
prediction. Moreover, the exact results are in very good agreement with the
pair MF results. Those for the ladder are in fact almost indistinguishable from
those of the chain, while those for the lattice are slightly closer to the pair
MF result due to the larger connectivity, in agreement with the perturbative
corrections of Eq.\ (\ref{rhokp}) (which can again predict the tails of
$S(\rho_{12})$ in the parity preserving phases). Conventional MF results, not
shown, are similar to those of Figs.\ \ref{f3}--\ref{f4}. The concurrence
$C(\rho_{12})$ remains close in the three systems also for higher values of the
total $\alpha$, as seen in the bottom panel.

\subsection{XYZ coupling}
Let us now examine the effects of an additional $J_z$ coupling  in (\ref{H}), i.e.,
\begin{equation}
H=\sum_{i=1}^n B (s^z_{2i-1}+s^z_{2i})-\sum_{\mu=x,y,z}J_\mu(s_{2i-1}^\mu
 s_{2i}^\mu+\alpha_\mu s_{2i}^\mu s_{2i+1}^\mu)\,. \label{H4}\end{equation}
As is well known, this model is no longer analytically solvable in the general
anisotropic case (the added term does not lead to a quadratic fermionic
operator in the Jordan-Wigner fermionization). We again assume $J_x>0$ and
$|J_y|<J_x$, with a common anisotropy $\alpha_\mu=\alpha>0$.

For small values of $J_z$, the phase diagram of Fig.\ \ref{f2} remains
essentially valid, with adequate shifts in the critical values of the field and
$\alpha$. At the conventional MF level, Eq.\ (\ref{thmf}) applies with
$B_c^\alpha$ replaced by the critical field
\[B_c^{\alpha z}=(J_x-J_z)(1+\alpha)s\,,\]
with no parity breaking phase if $J_z>J_x$. And a uniform factorizing field
still exists for common anisotropy if $J_z<J_y$, given by
\begin{equation}B_s^{\alpha z}=\sqrt{(J_x-J_z)(J_y-J_z)}(1+\alpha)s\,.\label{Bsz}\end{equation}
For $B=B_s^{\alpha z}$ the uniform parity breaking MF state
(\ref{mf1})--(\ref{0mf})  becomes again an exact degenerate GS \cite{CRM.10},
with $\cos\theta=\sqrt{\frac{J_y-J_z}{J_x-J_z}}$ (and $\theta=\pm |\theta|$).
If $J_z>J_x>J_y$, a factorized eigenstate still exists at $B=B_s^{\alpha z}$,
but will not be a GS \cite{CRM.10}.

At the pair MF level, we may still use the same state
(\ref{pst}), which leads to
\begin{equation}
\langle 0^p_h|H|0^p_h\rangle=\langle 0^p_h|H_{xy}|0^p_h\rangle-
\frac{n}{4}J_z[\cos\theta+\alpha\cos^2\phi\cos^4{\textstyle\frac{\theta}{2}}]\,,\label{Hmz}
\end{equation}
where $\langle 0^p_h|H_{xy}|0^p_h\rangle$ denotes Eq.\ (\ref{Ep}). Hence,
Eqs.\ (\ref{ss}) are to be replaced by
\begin{subequations}\label{ssz}
\begin{eqnarray}&&
\;\;\theta=0,\;\tan\phi=\frac{J_-}{B+\frac{1}{2}\alpha J_z\cos\phi}\,,\;\;
 B\geq B_{c2}^{\alpha z}\,,\label{aa}\\
&&{\left\{\begin{array}{l}\cos \theta=\frac{2(B\cos\phi+
J_-\sin\phi-J_+)+J_z(1+\frac{1}{2}\alpha\cos^2\phi)}
{\alpha(J_x(1+\sin\phi)-\frac{1}{2}J_z\cos^2\phi)}\\
\tan\phi=\frac{J_- +\frac{1}{4}\alpha J_x(1-\cos\theta)}
{B+\frac{1}{4}\alpha J_z\cos\phi(1+\cos\theta)}
\end{array}\right.
}\!\!\!\!\!\!\!\!\!
B_{c1}^{\alpha z}< B< B_{c2}^{\alpha z}\nonumber\\
&&\label{bb}\\
&&\;\;\theta=\pi\;(\phi\;{\rm arbitrary})\,,\;\;B\leq B_{c1}^{\alpha z}\,\label{cc}
\end{eqnarray}
\end{subequations}
where the critical fields depend now on $J_z$. The first critical field, which
delimits the maximally entangled dimerized phase, has still a simple exact
expression,  given by
\begin{eqnarray}
B_{c1}^{\alpha z}&=&{\textstyle\frac{1}{2}\sqrt{(J_x-J_z)(J_y-J_z-2\alpha J_x)}}\,.\label{Bc1z}
\end{eqnarray}
Eq.\ (\ref{Bc1z}) implies that for $J_z<J_y$, this dimerized phase will exist
for $\alpha<\alpha_{cz}$, with
\begin{equation}
\alpha_{cz}=\frac{J_y-J_z}{2J_x}\,.\label{alz}
\end{equation}
If $\alpha>\alpha_{cz}$ (or $J_z>J_y$)  parity will
be broken for all $B<B_{c2}^{\alpha z}$. $B_{c2}^{\alpha z}$ will also vanish for
sufficiently large $J_z$.

\begin{figure}[t]
\vspace*{0cm}

\centerline{\hspace*{-0.2cm}\scalebox{.7}{\includegraphics{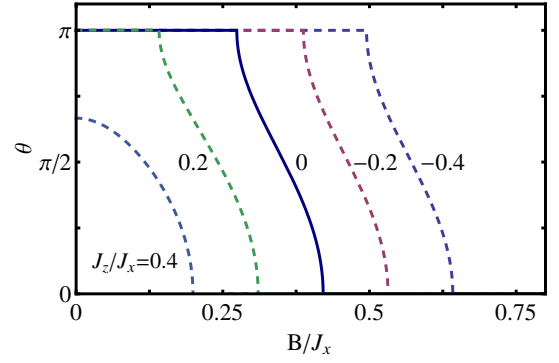}}}
\caption{(Color online) The angle $\theta$ of the pair MF approach for the
$XYZ$ Hamiltonian (\ref{H4}), as a function of the transverse field for
different values of $J_z/J_x$. The dimerized phase corresponds to $\theta=\pi$,
the partially aligned phase to $\theta=0$ and the parity breaking phase to
$0<\theta<\pi$. We have set $J_y/J_x=1/2$ and $\alpha=0.1$. A positive
(negative) $J_z$ in (\ref{H4}) unfavors (favors) the dimerized phase, which
will exist for $J_z<J_y-2\alpha J_x$ (Eq.\  (\ref{alz})).}
 \label{f8}
\end{figure}

A positive $J_z$ in Eq.\ (\ref{H4}) obviously increases the energy of the
dimerized state ($\theta=\pi$ in Eq.\ (\ref{Hmz})). Hence, its effect will be
to decrease the critical fields, narrowing the dimerized phase as appreciated
in Fig.\ \ref{f8}.  This phase will in fact disappear for $J_z>J_y-2\alpha J_x$
(Eq.\ (\ref{Bc1z})), as also seen in Fig.\ \ref{f8}. On the other hand, a
negative $J_z$ has the opposite effect, lowering the energy of the dimerized
state and increasing $B_{c1}^{\alpha z}$, favoring dimerization.  This picture
will remain valid for sufficiently weak longer range $XYZ$ couplings, employing
the substitutions (\ref{rep}) or (\ref{alxy}).

\begin{figure}[t]
\centerline{\hspace*{-0.2cm}\scalebox{.7}{\includegraphics{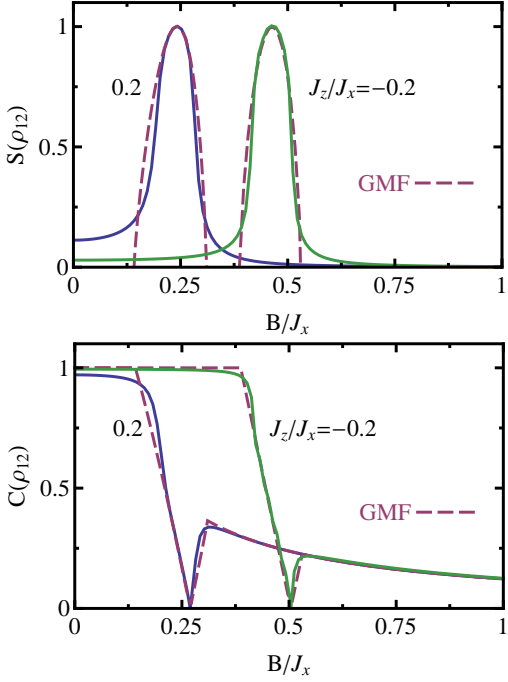}}}
\caption{(Color online) GS results for the $XYZ$ chain of Eq.\ (\ref{H4}). The
entanglement entropy $S(\rho_{12})$ (top) of strongly coupled pairs with the
rest of the chain and their concurrence $C(\rho_{12})$ (bottom) are plotted
for increasing fields at $\alpha=0.1$ for $J_z=\pm 0.2 J_x$. Exact results
(solid lines) are again in agreement with those of the pair MF (GMF, dashed
lines), which predicts a peak of $S(\rho_{12})$ in a displaced (with respect to
that for $J_z=0$) parity breaking sector, and a lower (higher) critical field
for the dimerized phase if $J_z>0$ ($J_z<0$). The concurrence vanishes at the
factorizing field (\ref{Bsz}). } \label{f9}
\end{figure}

Results for a finite cyclic $XYZ$ chain are depicted in Fig.\ \ref{f9}. Exact
results were again computed by diagonalization for $n=16$ spins.  It is
verified  that the pair MF predictions are fully confirmed. The addition of a
small $J_z$ coupling essentially shifts the results of the $XY$ chain, in
agreement with Eqs.\ (\ref{Bsz}) and (\ref{Bc1z}). As previously stated, a
reduced (extended) dimerized phase is obtained if $J_z>0$ ($J_z<0$), together
with a displaced parity breaking phase, which is still clearly visible through
the peak in the dimer entanglement entropy  $S(\rho_{12})$ with the rest of the
chain. There is again a good  agreement with the pair MF results, which can
also be improved by adding the corrections of Eq.\ (\ref{rhokp}). For strong
fields $B\gg B_{c2}^{\alpha z}$, we mention that the final effect is the
replacement $B\rightarrow B_{\rm eff}=B+\frac{1}{4}\alpha J_z$ (Eq.\
(\ref{aa})), with $\phi\approx J_-/B_{\rm eff}$ for $B\gg B_{c2}^{\alpha z}$.

\section{Conclusions\label{IV}}
We have investigated a general self-consistent variational MF approximation,
based on the selection of an arbitrary subset of operators for representing the
system density matrix, and its capability for describing entanglement in the GS
of composite systems. While retaining the conceptual simplicity of the
conventional MF, the generalization allows to significantly improve it by
considering composite cells, such that couplings within the cell are treated
exactly. The approach is then specially suitable for systems where a partition
in composite cells with strong internal couplings but weak cell-cell couplings
is feasible, although it is  not limited to this case.

In the dimerized systems considered, the approach naturally leads to a pair MF
approximation which is still analytic and simple, but which goes well beyond
the plain single spin MF. Its phase diagram clearly identifies a dimerized
phase for weak fields, together with a parity breaking phase in a transitional
region between the latter and the strong field regime. The approach is thus
able to accurately describe the entanglement of strongly coupled pairs, with
parity breaking emerging as a signature of a non-negligible entanglement
between these pairs and rest of system in the exact definite parity GS. With
the addition of simple perturbative corrections, it is also possible to predict
the concurrence of weakly coupled pairs and to improve the entanglement
predictions, as well as to describe the main features of the energy spectrum.

The generalized MF can be used as starting point for implementing more
sophisticated techniques. It is also directly applicable at finite
temperatures, higher spins, etc. These aspects and their application to more
complex systems are currently under investigation.

The authors acknowledge support from CONICET (AB, NC, JMM), and CIC
(RR) of Argentina.

\appendix*
\section{Exact solution of the cyclic dimer chain}
By means of the Jordan-Wigner transformation \cite{LSM.61}, and for a {\it
fixed} value $P=\pm $ of the global $S_z$-parity $P_z$ (Eq.\ (\ref{Pz})), we
may exactly rewrite the dimerized Hamiltonian (\ref{H1}) as a quadratic form in
standard fermion creation and annihilation operators $c^\dagger_j$, $c_j$,
which in terms of the spin operators read
\begin{eqnarray}
c^\dagger_{j}=s^+_j\exp[-i\pi\sum_{k=1}^{j-1}s^+_{k}s^{-}_{k}],
\end{eqnarray}
where $s^{\pm}_j=s^{x}_j\pm is^{y}_j$. These operators fulfill the fermionic
anticommutation relations $[c_j,c^\dagger_k]_+=\delta_{jk}, [c_j,c_k]_+= 0$.
The corresponding inverse transformation is
\begin{eqnarray}
s^+_j=c^\dagger_{j}\exp[i\pi\sum_{k=1}^{j-1}c^\dagger_{k}c_{k}].
\end{eqnarray}
We then obtain, setting $J_{\pm}=\frac{J_x\pm J_y}{4}$,
\begin{equation}
H^{P}=\!\sum_{j=1}^{2n} B(c^\dagger_jc_j-\half)-
\eta^{P}_j r_j(J_{+}c^\dagger_j c_{j+1}
+J_{-}c^\dagger_j c^\dagger_{j+1}+h.c.)\label{hf1}
\end{equation}
where $r_j=\left\{^{1\;(j\;{\rm odd})}_{\alpha\;(j\;{\rm even})}\right.$ and
$\eta^+_j=1-2\delta_{j,2n}$,  $\eta^-_j=1$  in the cyclic case. Through
separate parity dependent discrete Fourier transforms for even and odd sites,
\[\left(\begin{array}{c}c^{\dagger}_{2j-1}\\c^\dagger_{2j}\end{array}\right)=
\frac{1}{\sqrt{n}}\sum_{k\in K_{P}}
e^{-i2\pi k j/n}\left(\begin{array}{c}{c'}^\dagger_{k-}\\ {c'}^\dagger_{k+}
\end{array}\right)\,, \]
where $K_+=\{\half,\ldots,n-\half\}$,
$K_-=\{0,\ldots,n-1\}$, we may rewrite (\ref{hf1}) as
 \cite{CRM.10}
\begin{eqnarray}H^P&=&\sum_{k\in K_{P}}[\sum_{\sigma=\pm}
B({c'}^\dagger_{k\sigma}{c'}_{k\sigma}-\half)\nonumber\\&&-
(J_+^k{c'}^\dagger_{k-}{c'}_{k+}
+J_-^k{c'}^\dagger_{k-}{c'}^{\dagger}_{-k+}+h.c.)]\nonumber\\
&=&\sum_{k\in K_{P}}\sum_{\nu=\pm}
\lambda_k^\nu (a^\dagger_{k\nu}a_{k\nu}-\half)\,,
 \label{Hdf}\end{eqnarray}
where $J_{\pm}^k=J_{\pm}(1\pm\alpha e^{-i2\pi k/n})$ and $-k\equiv n-k$.
The final diagonal form
(\ref{Hdf}) is obtained by means of a Bogoliubov transformation
${c'}^\dagger_{k\sigma}=\sum_{\nu=\pm}U_{k\sigma}^\nu
a^\dagger_{k\nu}+V_{k\sigma}^\nu a_{-k\nu}$ determined through the
diagonalization of $4\times 4$ blocks
\begin{equation}{\cal H}_k=\left(\begin{array}{cccc}B&-J_+^k&0&-J_-^k\\
-\bar{J}_+^k&B&\bar{J}_-^k&0\\0&J_-^k&-B&J_+^k\\
-\bar{J}_-^k &0&\bar{J}_+^k&-B
 \end{array}\right)\,,\end{equation}
whose eigenvalues are $\pm\lambda_k^+$, $\pm\lambda_k^-$, with
\begin{equation}|\lambda_{k}^{\pm}|=\sqrt{\Delta\pm\sqrt{\Delta^2-
|B^2-(J_+^k+J_-^k)(\bar{J}_+^k-\bar{J}_-^k)|^2}} \label{4.8}
\end{equation} and
$\Delta=B^2+|J_+^k|^2+|J_-^k|^2$. Care should be taken to select the correct
signs of $\lambda_k^\pm$ in order that the vacuum of the operators $a_{k\nu}$
has the proper $S_z$-parity and represents the lowest state for this parity.

The spin correlations in the lowest states for each parity can then be obtained
from the ensuing basic fermionic contractions $f_{ij}=\langle c^\dagger_i
c_j\rangle-\half \delta_{ij}$, $g_{ij}=\langle c^\dagger_i c^\dagger_j\rangle$,
which can be directly obtained from the inverse Fourier transform of $\langle
{c'}^\dagger_{k\sigma}{c'}_{k\sigma'}\rangle=\sum_\nu V_{k\sigma}^\nu
\bar{V}_{k\sigma'}^\nu$, $\langle
{c'}^\dagger_{k\sigma}{c'}^\dagger_{-k\sigma'}\rangle=\sum_\nu  V_{k\sigma}^\nu
U_{-k\sigma'}^\nu$. We then obtain, through the use of Wick's theorem, $\langle
s^z_i\rangle=f_{ii}$, $\langle
s^z_is^z_j\rangle=f_{ii}f_{jj}-f^2_{ij}+g_{ij}^2$, and $\langle
s_{i}^+s_{j}^\mp\rangle=\frac{1}{4}[{\rm det}(A^+_{ij})\pm {\rm
det}(A^-_{ij})]$, where $A^\pm_{ij}$ are $(j-i)\times(j-i)$ matrices of
elements $2(f+g)_{i+p+^0_1,i+q+ ^1_0}$, with $p,q=0,\ldots,j-i-1$.

From Eq. (\ref{4.8}) it is seen that for real $B\neq 0$ and finite $n$,
$|\lambda_k^+|>0$ while  $\lambda_k^-$ vanishes just when $k=0$  and
\begin{equation}
 B=B_{c2}^{\rm ex}=\frac{1}{2}\sqrt{(\alpha J_x+J_y)(J_x+\alpha J_y)},\label{AB}
\end{equation}
or  $k=n/2$  and
\begin{equation}
 B=B_{c1}^{\rm ex}=\frac{1}{2}\sqrt{(J_y-\alpha J_x)(J_x-\alpha J_y)},\label{AA}
\end{equation}
remaining non-zero for other values of $k$. These critical fields coincide with
those of refs.\ \cite{Pr.75,GG.09} for the present situation. For $0\leq
\alpha\leq 1$ and $J_x>0$,  Eq.\ (\ref{AA}) is real only for $J_y\geq 0$ and
$\alpha \leq J_y/J_x$, while if $-J_x\leq J_y\leq 0$, Eq.\ (\ref{AB}) is real
for $\alpha\geq -J_y/J_x$. The pair MF critical fields (\ref{Bc1})--(\ref{Bc2})
correspond approximately to these fields
 and satisfy
\begin{equation}
B_{c1}\leq B_{c1}^{\rm ex}\leq B_s^\alpha\leq B_{c2}^{\rm ex}\leq B_{c2}
\,,\end{equation}
for $J_y\geq 0$, all approaching the factorizing field $B_s^0=\frac{\sqrt{J_x
J_y}}{2}$ for $\alpha\rightarrow0$ (where $B^{\rm ex}_{c1,c2}\approx B_s^0[1\mp
\frac{\alpha}{2}(\frac{J_x}{J_y}+\frac{J_y}{J_x})]$).

The fields (\ref{AB})--(\ref{AA}) enclose the interval where the finite chain
GS will be almost two-fold degenerate, i.e., where the lowest state with
positive $S_z$ parity will have nearly the same energy as the lowest state with
negative parity. Actually, starting at a field slightly above $B=B_{c1}^{\rm
ex}$, the exact GS of the finite chain will experience $n$  parity transitions
\cite{CRM.10,RCM.08} in the interval ($B_{c1}^{\rm ex}, B_{c2}^{\rm ex})$, with
the last one taking place exactly at the factorizing field $B_s^\alpha$.


\begin{thebibliography}{999}
\bibitem{Am.08} L.\ Amico et al, Rev.\ Mod.\ Phys.\ {\bf 80}, 517 (2008).
\bibitem{ECP.10}J.\ Eisert, M.\ Cramer, M.B.\ Plenio, Rev.\ Mod.\ Phys.\ {\bf 82}, 277 (2010).
\bibitem{NC.00} M.A.\ Nielsen, I.L.\ Chuang, {\em Quantum Computation and Quantum Information}
(Cambridge Univ. Press, Cambridge, UK, 2000).
\bibitem{HR.06} S.\ Haroche, J.M. Raimond {\em Exploring the Quantum}
(Oxford,  Univ. Press, Oxford, UK (2006).
\bibitem{ON.02} T. J. Osborne, M.A.\ Nielsen, Phys.\ Rev.\ A {\bf 66}, 032110
(2002); G.\ Vidal, J.I.\ Latorre, E.\ Rico, A.\ Kitaev, Phys.\ Rev.\ Lett.\
{\bf 90}, 227902 (2003).
\bibitem{RS.80}P.\ Ring, P.\ Schuck, {\em The Nuclear Many-body Problem}  (Springer, Berlin, 1980);
J.P.\ Blaizot, G.\ Ripka, {\em Quantum Theory of Finite Systems} (MIT Press, Cam.\, Ma., 1986).
\bibitem{DMRG.03} S.R.\ White, Phys.\ Rev.\ Lett.\ {\bf 69}, 2863 (1992);
U. Schollwock, Rev.\ Mod.\ Phys.\ {\bf 77},  259 (2003); Ann.\ of  Phys.\ {\bf 326}, 96 (2011).
\bibitem{OS.95} S.\ \"Ostlund, S.\ Rommer, Phys.\ Rev.\ Lett.\ {\bf 75} 3537 (1995);
S.\ Rommer, S.\ \"Ostlund, Phys.\ Rev.\ B {\bf 55} 2164 (1997).
\bibitem{VC.08} F.\ Verstraete, J.I.\ Cirac,
V.\ Murg, Adv.\ Phys.\  {\bf 57}, 143 (2008);
F. Verstraete, J.I.\ Cirac, Phys.\ Rev.\ B {\bf 73} 094423 (2006);
J.I.\ Cirac, F.\ Verstraete,  J.\ Phys.\ A {\bf 42}, 504004 (2009).
\bibitem{GV.08}G.\ Vidal, Phys.\ Rev.\ Lett.\ {\bf 99}, 220405 (2007);
{\bf 101}, 110501 (2008); A.J.\ Ferris, G.\ Vidal, Phys.\ Rev.\ B {\bf 85}, 165147 (2012).
\bibitem{LDA.88} S.\ Liang, B.\ Doucot and P.W.\ Anderson, Phys.\ Rev.\ Lett.\ {\bf 61}, 365 (1988).
\bibitem{LS.07} J.\ Lou, A.W.\ Sandvik, Phys.\ Rev.\ B {\bf 76}, 104432 (2007);
A.W.\ Sandvik, H.G.Evertz, Phys.\ Rev.\ B {\bf 82}, 024407 (2010).
\bibitem{VMC.01} W.M.C.\ Foulkes et al,
Rev.\ Mod.\ Phys.\ {\bf  73}, 33 (2001); R.J.\ Needs et al, J.\ Phys.\ Cond.\
Matter\ {\bf 22}, 023201 (2010).
\bibitem{RC.97} R.\ Rossignoli, N.\ Canosa, Phys.\ Lett.\ B {\bf 394}, 242 (1997);
R.\ Rossignoli, N.\ Canosa, P.\ Ring, Phys.\ Rev.\ Lett.\ {\bf 80}, 1853
(1998); N.\ Canosa, R.\ Rossignoli, Phys.\ Rev.\ B {\bf 62} 5886 (2000).
\bibitem{CRM.07}
N.\ Canosa, J.M.\ Matera, R.\ Rossignoli, Phys.\  Rev.\ A {\bf 76}, 022310
(2007); J.M.\ Matera, R.\ Rossignoli, N.\ Canosa, Phys.\ Rev.\ A {\bf 78},
042319 (2008); Phys.\ Rev.\ A {\bf 82}, 052332 (2010).
\bibitem{MC.04} S.\ Mukhopadhyay, I.\ Chatterjee, J.\  Magn.\ Magn.\ Mater.\
{\bf 270}, 247 (2004).
\bibitem{DJ.02} M.\ Dantziger et al, Phys.\ Rev.\ B {\bf 66} 094416 (2002);
I.\ Etxebarria, L.\ Elcoro, J.M.\ Perez-Mato, Phys.\ Rev.\ E {\bf 70}, 066133 (2004).
\bibitem{Y.09} D.\ Yamamoto, Phys.\ Rev.\ B {\bf 79}, 144427 (2009);
Y.Z.\ Ren, N.H.\ Tong, X.C.\ Xie, J.\ Phys: Cond.\ Matt.\ {\bf 26}, 115601 (2014).
\bibitem{BPW.35} H.A.\ Bethe, Proc.\ R.\ Soc.\ London, Ser.\ A {\bf 150}, 552 (1935);
 R.E.\ Peierls, Proc.\ Camb.\ Phil.\ Soc.\ {\bf 32}, 477 (1936);
 P.R.\ Weiss Phys.\ Rev.\ {\bf 74}, 1493 (1948);
T.\ Oguchi, Prog.\ Theor.\ Phys.\ {\bf 13}, 148 (1955).
\bibitem{RP.90} R.\ Rossignoli, A.\ Plastino, Phys.\ Rev.\  A {\bf  42}, 2065 (1990);
Phys.\ Rev.\ C {\bf  40}, 1798 (1989);
R. Rossignoli, A.\ Plastino, H.G.\ Miller, Phys.\ Rev.\ C {\bf 43}, 1599 (1991).
\bibitem{Pr.75} J.H.H.\ Perk, H.W.\ Capel, M.J. Zuilhof, Th.J.\ Siskens,
Phys.\ A {\bf 81}, 319 (1975);
Th.J.\ Siskens, H.W.\ Capel, J.H.H.\ Perk, Phys.\ Lett.\ A {\bf 53}, 21 (1975).
 \bibitem{Pr.77} J.H.H.\ Perk,  H.W.\ Capel, Th.J.\ Siskens, Phys.\ A {\bf 89}, 304 (1977);
 J.H.H.\ Perk,
H.W.\ Capel, Phys.\ A {\bf 92}, 163 (1978).
\bibitem{Pr.09} J.H.H.\ Perk, H.\ Au-Yang, J.\ Stat.\ Phys.\ {\bf 135}, 599 (2009).
\bibitem{FM.07} E.I.\ Kutznetsova, E.B.\ Fel'dman, JETP.\ Lett.\ {\bf 102}, 882 (2006);
E.B.\ Fel'dman, M.G.\ Rudavets, JETP.\ Lett.\ {\bf 81}, 47 (2005).
S.D. Doronin et al, JETP Lett.\ 85, 519 (2007).
\bibitem{GG.09}G.L.\ Giorgi, Phys.\ Rev.\ B {\bf 79}, 060405(R) (2009); {\bf 80},
019901(E)(2009).
\bibitem{CRM.10} N.\ Canosa, R.\ Rossignoli, J.M.\ Matera, Phys.\ Rev.\ B {\bf 81},
054415 (2010).
\bibitem{LSM.61} E.\ Lieb, T.\  Schultz, D.\ Mattis, Ann.\ Phys.\ {\bf 16}
407 (1961).
\bibitem{MG.69} C.K.\ Majundar, D.K.\ Gosh, J.\ Math.\ Phys.\ {\bf 10}, 1388 (1969);
{\it ibid} {\bf 10}, 1399 (1969); B.S.\ Shastry and B.\ Sutherland,
Phys.\ Rev.\ Lett.\ {\bf 47}, 964  (1981).
\bibitem{S.05}H.J.\ Schmidt, J.\ Phys.\ A {\bf 38} 2123 (2005).
\bibitem{KSV.07} D.\ Kaszlikowski, W.\ Son, V.\ Vedral, Phys.\ Rev.\ A {\bf 76}, 054302
(2007).
\bibitem{HXG.08} M.-G.\ Hu, K.\ Xue, M.-L.\ Ge, Phys.\ Rev.\ A {\bf 78}, 052324
(2008).
\bibitem{HH.11} J.\ Sirker et al Phys.\ Rev.\ Lett. {\bf 101} 157204 (2008);
 A.Herzog et al, Phys.\ Rev.\ B {\bf 84} 134428 (2011).
\bibitem{Mn.14} P.\ Merchant et al, Nature Phys.\ {\bf 10}, 373 (2014).
\bibitem{ZZ.14}  T.\ Ramos et al, arXiv:1408.4357 [quant-ph] (2014);
A.W.\ Glaetzle et al, arXiv:1410.3388 [quant-ph] (2014); I.\ Bloch,
 J.\ Dalibard, W.\ Zwerger, Rev.\ Mod.\ Phys.\ {\bf 80}, 885 (2006).
\bibitem{Wo.97} S.\ Hill and  W.K.\ Wootters, Phys.\ Rev.\ Lett.\ {\bf 78}, 5022 (1997);
W.K.\ Wootters, Phys.\ Rev.\ Lett.\ {\bf 80}, 2245 (1998).
\bibitem{KP.06} A.\ Kitaev, and  J.\ Preskill, Phys.\ Rev.\ Lett.\ {\bf 96}, 110404 (2006);
M.\ Levin, X.G.\ Wen, Phys.\ Rev.\ Lett.\ {\bf 96}, 110405 (2006).
\bibitem{GAI.09}
 S.M.\ Giampaolo, G.\ Adesso, F.\ Illuminati, Phys.\ Rev.\ B {\bf 79}, 224434 (2009);
 Phys. Rev. Lett. {\bf  100}, 197201 (2008).
\bibitem{RCM.08} R. Rossignoli, N. Canosa, J.M.\ Matera, Phys.\ Rev.\ A {\bf 77}, 052322 (2008);
Phys.\ Rev.\ A {\bf 80}, 062325 (2009).
\bibitem{KTM.82} J.\ Kurmann, H.\ Thomas, G.\ M\"uller, Phys.\ {\bf A} 112, 235 (1982).
\bibitem{WW.89} R.F.\ Werner, Phys.\ Rev.\ A {\bf 40}, 4277 (1989).
\end{thebibliography}
\end{document}